\RequirePackage{fix-cm}
\documentclass[twocolumn,epjc3]{svjour3}  
\smartqed  
\RequirePackage{graphicx}
\usepackage{amsmath}
\usepackage{cuted} 

\journalname{Eur. Phys. J. C}
\begin{document}

\title{Enhanced tensor non-Gaussianities in presence of a source
}


\author{Abhishek Naskar\thanksref{e1,addr1}
        \and
        Supratik Pal\thanksref{e2,addr1} 
}

\thankstext{e1}{e-mail: abhiatrkmrc@gmail.com}
\thankstext{e2}{e-mail: supratik@isical.ac.in}

\institute{Physics and Applied Mathematics Unit, Indian Statistical Institute, Kolkata, 700108, India \label{addr1}
}

\date{Received: date / Accepted: date}

\maketitle

\begin{abstract}
We address the possibility of having an enhanced signal for tensor non-Gaussianities in presence of a source,
as a signature of Primordial Gravitational Waves.
We employ  a nearly model-independent framework based on Effective Field Theory of inflation
and compute tensor non-Gaussianities therefrom sourced by particle production during (p)reheating to arrive at an enhanced signal strength.
We obtain the non-linearity parameters and also find that squeezed limit bispectra are more enhanced than equilateral limit.
\end{abstract}

\section{Introduction}

Even after the profound advancement in the Cosmic Microwave Background (CMB) observations for nearly two decades,
Primordial Gravitational Waves (PGW) the so-called tensor modes of perturbations still remain as the holy grail of  early universe cosmology.
The latest bound on the
 amplitude of two-point correlation function of tensor modes {\it i.e} tensor-to-scalar ratio
   is $r<0.064$  from Planck 2018 data \cite{Akrami:2018odb}. All it gives us is an impression that the signal strength of
 power spectrum for  PGW, if exists, would be really tiny, making it a daunting task for next-generation CMB missions to detect it some day. 
 Despite this,  from theoretical point of view,  PGW encodes crucial
information about early universe cosmology. 
PGW generated due to vacuum fluctuations during inflation is directly related to 
inflationary energy scale. 
In absence of any conclusive evidence of two-point function for PGW until now, the community got curious about the three-point function
that reflects the non-Gaussian features of PGW, primarily because it has potential to serve as an additional probe of PGW.
Over the last few years  there has been some theoretical progress
 in this direction.
   In \cite{mald1,mald2} the
 three-point function for tensor modes is calculated for general single field
slow roll inflationary models. This analysis is further generalized 
in \cite{yamaguchi,Naskar:2018rmu}.
  For a recent review the reader can refer to \cite{Shiraishi:2019yux}.
   These
 analysis are for tensor modes generated by vacuum fluctuations.
  However,     
it has been pointed out in a previous article by the present authors  \cite{Naskar:2018rmu} in a model-independent framework 
based on EFT of inflation, and also 
by others following particular models,
that the amplitude of bispectrum generated by vacuum fluctuations is generically small. 
 
  Apart from vacuum fluctuations, PGW can also be 
generated by some sources that may be present during the  early epoch.
While some of these sources can affect the powerspectrum of PGW non-trivially,
 one can also investigate for 
non-Gaussian features of PGW which has different
momentum dependence for different sources and hence can distinguish among different sources and vacuum.
Of late this revelation has served as a strong motivation to explore non-Gaussian statistics of PGW from possible sources. 
 Subsequently,  the possibilities of producing comparatively large
  signal using different sources  have been investigated to some extent, for example, 
 using axion as a source \cite{Namba:2015gja,aniket}, or 
 using extra spin particles during inflation \cite{Dimastrogiovanni:2018gkl}.
    
The current observations are unable to detect any significant signal of tensor non-Gaussianities. 
Latest constraints
 on the amplitude of three-point function with $1\sigma$ error are
 $f_{NL}^{T} = 600 \pm 1500$   from
 WMAP  \cite{Shiraishi:2014ila} and $f_{NL}^{T} = 800 \pm 1100$ from Planck 2018  \cite{Akrami:2019izv}
 for equilateral momentum configuration and on the amplitude for tensor-scalar-scalar three point
 function are $f_{NL}^{TSS}=84\pm49$ at $68\%$C.L. \cite{Shiraishi:2017yrq}.
Nonetheless,  the methodology for bispectrum estimation is established
by adding B-mode polarization information \cite{Shiraishi:2019yux}. Upcoming CMB mission 
LiteBIRD \cite{Matsumura:2013aja,Suzuki:2018cuy} targets to improve the results by three 
orders of magnitude. CMB-S4 \cite{Abazajian:2016yjj} may improve the tensor-scalar-scalar
cross correlation result by an order of magnitude. The dedicated gravitational waves detector LISA \cite{Bartolo:2018qqn} can
directly probe the bispectrum of gravitational waves. Future missions like Advanced LIGO \cite{TheLIGOScientific:2014jea}, BBO \cite{Crowder:2005nr}
will work with improved sensitivity towards the detection of tensor non-Gaussianity.
 So it is important to do a 
theoretical analysis on generic aspects of tensor non-Gaussian statistics and interpret the constraints in the light of upcoming
observations.

    In this article we intend to take up our previous model-independent analysis  \cite{Naskar:2018rmu} based on EFT of inflation 
    and extend it to possible sources. 
     We want to explore if it is possible to enhance the bispectrum of PGW 
    due to (p)reheating process. To this end we will make use of the 
    EFT of inflation \cite{crem2} and EFT of (p)reheating \cite{Giblin:2017qjp}. As in the case of our previous analysis \cite{Naskar:2018rmu}, the present analysis depends solely on the EFT parameters
    and different choice of these parameters leads to different models.
    In particular, we would
    be interested in proposing expressions for non-linearity parameter $f_{NL}$ from the 
    model independent framework of EFT.


\section{EFT, Graviton Lagrangian and (P)reheating} \label{sec:eft}
	As mentioned, since our intention is to analyze the scenario in  a more or less  model independent 
	framework, we make use of the EFT of inflation following our previous analysis  \cite{Naskar:2018rmu},
	that was originally  developed in \cite{crem2,weinberg}. 
	 In this approach, the inflaton field $\phi$ is a scalar under all diffeomorphisms but $\delta \phi$ breaks the time 
	 diffeomorphism. Using this symmetry of the system and unitary gauge where $\delta \phi=0$, the  Lagrangian 
	 can be written as  \cite{crem2}

\begin{multline}\label{eq1}
	\mathcal{S}=\int d^4x \sqrt{-g}\left[\frac{1}{2}M_{pl}^2R-\Lambda(t)-c(t)g^{00} \right. \\
           \left. 
	 +\frac{1}{2}M_2(t)^4(g^{00}+1)^2-\frac{\bar{M}_1(t)^3}{2}(g^{00}+1)\delta K_{\mu}^{\mu}\right. \\
           \left. 
           -\frac{\bar{M}_2(t)^2}{2}\delta K_{\mu}^{\mu2} 
           -\frac{\bar{M}_3(t)^2}{2}\delta K_{\mu}^{\nu} \delta K_{\nu}^{\mu} +  \frac{M_3(t)^4}{3!}(g^{00}+1)^3\right. \\
           \left. -\frac{\bar{M}_4(t)^3}{3!}(g^{00}+1)^2\delta K_{\mu}^{\mu}
            -\frac{\bar{M}_5(t)^2}{3!}(g^{00}+1)\delta K_{\mu}^{\mu 2}\right.\\
            \left.-\frac{\bar{M}_6(t)^2}{3!}(g^{00}+1)
            \delta K_{\mu}^{\nu } \delta K_{\nu}^{\mu}
            -\frac{\bar{M}_7(t)}{3!}\delta K_{\mu}^{\mu 3}\right. \\
           \left. 
            -\frac{\bar{M}_8(t)}{3!}\delta K_{\mu}^{\mu} \delta K_{\nu}^{\rho} \delta K_{\rho}^{\nu}-
            \frac{\bar{M}_9(t)}{3!}\delta K_{\mu}^{\nu} \delta K_{\nu}^{\rho} \delta K_{\rho}^{\mu}+....\right] .
\end{multline}

The dots at the end of the Lagrangian represent higher order fluctuation terms. 
As pointed out in  \cite{crem2}, this is purely gravitational Lagrangian where $R$ is the Einstein 
curvature term, $g^{00}$ is the time-time component of the metric tensor, $K_{\mu}^{\nu}$
is the extrinsic curvature, $\Lambda(t)$, $c(t)$, $M_i$ and
$\bar{M}_i$ are the parameters of the theory where parameters $\Lambda(t)$ and $c(t)$ can be fixed
by background evolution. The parameters $M_i$ and $\bar{M}_i$ can in general be time dependent but in 
our analysis we consider them as constants as the time dependence of these parameters is slow roll 
suppressed. In \eqref{eq1}
the scalar perturbation is not explicit but can be reintroduced using 
	 $St\ddot{u}ckleberg$ trick. 

 
    In Unitary gauge the perturbed metric can be written as, $g_{ij}(t,x)=a^2(t)[(1+2\zeta(t,x))\delta_{ij}+\gamma_{ij}(t,x)]$,
  where $a(t)$ is scale factor, $\zeta(t,x)$ is scalar perturbation and $\gamma_{ij}(t,x)$ 
  is tensor perturbation which
 is transverse and traceless satisfying,
  $\gamma_{ii}=0$ and $ \partial_{j} \gamma_{ij}=0$.
 In terms of $\gamma_{ij}$ the  Lagrangian \eqref{eq1} takes the form
  \begin{multline}\label{eq2}
  S_{3}^{T}=\int d^4x \sqrt{-g}\left[\frac{M_{pl}^2}{8}\left(\dot{\gamma}_{ij}^2-
  \frac{(\partial_k \gamma_{ij})^2}{a^2}\right)-\frac{\bar{M}_{3}^{2}}{8}\dot{\gamma}_{ij}^2 \right. \\
           \left.  -  \frac{M_{pl}^2}{8}\left(2 \gamma_{ik} \gamma_{jl}-\gamma_{ij}\gamma_{kl} \right)
\frac{\partial_k \partial_l \gamma_{ij}}{a^2}-\frac{\bar{M}_9}{3!} \dot{\gamma}_{ij}  \dot{\gamma}_{jk}  
\dot{\gamma}_{ki}\right],
  \end{multline}
 where a dot on the operators denotes derivative with respect to time. The propagation speed
  of tensor fluctuation gets modified as $ c_{\gamma}^2= \frac{M_{pl}^2}{M_{pl}^2-\bar{M}_3^2}$ due to the
 presence of $\bar{M_3}$ parameter.

  Eq \eqref{eq2} is the most general third order Lagrangian for single field inflation. It has been shown that the
 term proportional to  $\bar{M_9}$ along with the Einstein term contribute to tensor 
 bispectrum  \cite{Naskar:2018rmu}.
For our present investigation, our intention is to add, on top of this, the EFT of  (p)reheating that was developed in  \cite{Giblin:2017qjp}.
Here, apart from
  the inflaton fluctuation, one more  degree of freedom is considered. This approach
  also assumes that the background breaks the time diffeomorphism spontaneously 
  and the construction of the Lagrangian is similar as \cite{crem2}. For (p)reheat field $\chi$ 
  it can be written as,
  \begin{multline}\label{eq3}
  \mathcal{S}_{\chi}=\int d^4 x \sqrt{-g}\left[-\frac{\alpha_1(t)}{2}g^{\mu \nu}
   \partial_{\mu}\chi \partial_{\nu}\chi+\frac{\alpha_2(t)}{2}(\partial^0\chi)^2 \right. \\
           \left. 
     -\frac{\alpha_3(t)}{2}\chi^2 + \alpha_4 \chi \partial^0 \chi \right].
  \end{multline}
Here $\alpha_i$'s are parameters of the theory. With time repara\- metrization invariance, parameter 
$\alpha_4$  has been set to zero \cite{Giblin:2017qjp}.
 Note that the (p)reheat particles also have non-trivial propagation speed
  \begin{equation}
  c_{\chi}^2=\frac{\alpha_1}{\alpha_1+\alpha_2}.
  \end{equation}
  In our analysis we consider $\alpha_1$ and $\alpha_2$ to be time independent and 
  hence the propagation speed is also time independent.


\section{Two-point correlation function}  \label{sec:2pt}
 With (p)reheating particles as source with energy-momen\- tum tensor $T_{ab}(x,t)$ the equation of motion
 for $\gamma_{ij}(x, t)$ is given by,
\begin{equation}\label{eq4}
\gamma_{ij}^{''}(x, \tau)-2\frac{a'}{a}\gamma_{ij}'(x, \tau)+c_{\gamma}^2\Delta\gamma_{ij}
(x, \tau)=\frac{2}{M_p^2}\Pi_{ij}^{ab}T_{ab}(x, \tau).
\end{equation}
Here $'$ denotes derivative with respect to conformal time $\tau$, and 
 $\Pi_{ij}^{ab}$ is the transverse traceless projection tensor. Written explicitly,
\begin{equation}
\Pi_{ij}^{ab}=\Pi_{i}^{a}\Pi_{j}^{b}-\frac{1}{2}\Pi_{ij}\Pi^{ab}, ~~{\rm with}~~  \Pi_{ij}=\delta_{ij}-\frac{\partial_i \partial_j}{\Delta}.
\end{equation}
So the transverse traceless part of energy momentum tensor becomes

\begin{equation}\label{ttem}
\Pi_{ij}^{ab}T_{ab}=-\alpha_1 \Pi_{ij}^{ab}\partial_a \chi \partial_b \chi.
\end{equation}

Taking Fourier transform the solution for Eq \eqref{eq4} can be obtained by  
Green's function method, 
\begin{equation}
\gamma_{ij}(k,\tau)=\frac{2}{M_p^2}\int d\tau' G_k(\tau,\tau')\Pi_{ij}^{ab}T_{ab}(k,\tau'),
\end{equation}
where the expression for Green's function $G_k(\tau,\tau')$ is given by,

\begin{multline}\label{eq5}
G_k(\tau,\tau')=\frac{1}{c_{\gamma}^3k^3 \tau'^2}\left[(1+c_{\gamma}^2k^2 \tau \tau') \sin c_{\gamma}k(\tau-\tau')+\right. \\
\left. c_{\gamma}k(\tau'-\tau) \cos c_{\gamma}k (\tau-\tau')\right] \Theta(\tau-\tau')].
\end{multline}
 It is worthwhile to mention that
in \eqref{eq5} the non trivial propagation speed of tensor fluctuation plays a crucial 
role in determining the
 Green's function and hence the powerspectrum. This will be obvious from the following analysis.
In what follows we  employ the method of \cite{Cook} to calculate
 the two-point correlation function for our setup of nontrivial contribution from the EFT parameters.

Using this Green's function the power spectrum for tensor modes sourced by (p)reheat field turns out to be
\begin{multline}\label{eq6}
\left\langle \gamma_{ij}(k,\tau)\gamma^{ij}(k,\tau')\right\rangle=\frac{\alpha_1^2}{2\pi^3 M_p^4}
 \int \frac{d \tau'}{a(\tau')^2}G_k(\tau,\tau') \\
 \times \int \frac{d \tau''}{a(\tau'')^2}G_k(\tau,\tau'')
  \Pi_{ij}^{ab}(k)\Pi_{ij}^{cd}(k')\\ 
  \times \int d^3p d^3 p' p_a(k_b-p_b) p_c' (k_d'-p_d')\\
\times  \left \langle \chi(p,\tau')\chi(k-p,\tau')\chi(p',\tau'')\chi(k'-p',\tau'')\right\rangle.
\end{multline}

In order to evaluate the correlation functions we need to analyze the dynamics
of $\chi$ particles. Varying  \eqref{eq3} with respect to $\chi$ one arrives at the following equation of parametric oscillator
\begin{equation}\label{eq6}
\chi_c ''(k,\tau) + \omega^2 (k,\tau) \chi_c (k,\tau)=0,
\end{equation}
where, $\chi_c=a \chi (\alpha_1+\alpha_2)$ and the frequency of the oscillator is given by
\begin{equation}
\omega^2 (k,\tau)=k^2 c_{\chi}^2+a^2 (\tau)\frac{\alpha_3(t)}{\alpha_1+\alpha_2}-\frac{a''}{a}.
\end{equation} 
This clearly shows the nontrivial modifications to the frequency that arises due to the EFT of (p)reheating.

Consequently,  the solution for \eqref{eq6} becomes
\begin{equation}\label{eq7}
\chi_c=\frac{1}{\sqrt{2 \omega}}\left(\alpha(k,\tau) e^{-i\int^\tau \omega}+\beta(k,\tau)e^{i\int^\tau \omega}\right),
\end{equation}
where $\alpha$ and $\beta$ are the Bogolyubov coefficients.

To proceed further, we need to find explicit time dependence of  $\omega(k,\tau)$ i,e we need to
find the functional form of $\frac{\alpha_3(t)}{\alpha_1+\alpha_2}$. In order to do that we have to 
remember that there are two important energy scales in the theory: 
the cosmological time $H^{-1}$, $H$ being the Hubble parameter and
the time scale associated with the frequency of oscillations ($\omega_{osc}$) of inflaton at 
the end of inflation. This corresponds to a hierarchy of scales \cite{Giblin:2017qjp}.
At high energies $E>\omega_{osc}>H$ the time translation is unbroken. When $E<\omega_{osc}$  the 
time translation symmetry is broken as discrete symmetry and at even lower energy $E<H<\omega_{osc}$ 
cosmological expansion breaks time translation symmetry. As a consequence the background Hubble
parameter can be written as a sum of slowly time dependent function and an oscillatory function
\cite{Giblin:2017qjp,Behbahani:2011it},

\begin{equation}
H(t)= H_{sr}(t)+H_{osc}(t) P(\omega_{osc}t),
\end{equation}

where, $H_{sr}(t)$ and $H_{osc}(t)$ are slowly time dependent functions and $P(\omega_{osc}t)$ is 
some periodic function.
 Now the parameters of EFT of (p)reheating can be written as
a function of Hubble parameter and its derivatives \cite{Giblin:2017qjp} and hence will be periodic
in nature.
If we expand the periodic function $\frac{\alpha_3(t)}{\alpha_1+\alpha_2}$ with frequency 
$\omega_{\alpha_3}$ around its minimum $t_0$ then it can be written as,

\begin{equation}\label{alpha3}
\frac{\alpha_3(t)}{\alpha_1+\alpha_2} \propto \omega_{\alpha_3}^2 (t-t_0)^2+...
\end{equation}

In general the frequency $\omega_{\alpha_3}$ can be different than $\omega_{osc}$ 
and the dots represent higher 
order terms in the expansion. In our analysis we consider upto 
second order in time expansion. Physically the parameter $\alpha_3(t)$ describes the interaction 
between inflaton and $\chi$ particles. So our choice in \eqref{alpha3} can be written in an 
alternative way in terms of inflaton field,

\begin{equation}\label{alpha3-2}
\frac{\alpha_3(t)}{\alpha_1+\alpha_2}=\frac{g^2}{2}(\phi-\phi_0)^2,
\end{equation}

where, $\phi_0=\phi(t=t_0)$ and considering de-sitter background and with slow roll
approximation we can assume that, $\phi(t)= \dot{\phi_0}t$ where $\dot{\phi_0}$ is constant, so 
$t_0$ present in \eqref{alpha3} can be written as, $t_0=\frac{\phi_0}{\dot{\phi_0}}$. The parameter 
choice of \eqref{alpha3-2} is consistent with the background evolution and symmetry. With these
parameter choices of EFT of inflation and EFT of (p)reheating we are able to analyze the production
of PGW due to (p)reheating from a fairly general class of inflationary models and a 
class of (p)reheating
models where the propagation speed of produced particle is non-trivial and the interaction between 
inflaton and the (p)reheating particles is described by \eqref{alpha3} and \eqref{alpha3-2}.

With the parameter choice of \eqref{alpha3-2}, non-adiabatic condition leads to a constraint
 $g>>\frac{H^2}{\dot{\phi}_0}$, and with this constraint we can neglect the expansion of 
 universe and can consider $H$ as
 a constant in time \cite{Cook}. 
 With these approximations the Bogolyubov coefficients  turn out to be
 \begin{equation}
 \alpha(k,\tau>\tau_0)=\sqrt{1+e^{\frac{-c_{\chi}^2 k^2 H^2 \tau_0^2}{g \dot{\phi_0}}}} e^{i\alpha_k},
 \end{equation}
 and 
 \begin{equation}
 \beta(k,\tau>\tau_0)=i e^{\frac{-c_{\chi}^2 k^2 H^2 \tau_0^2}{2 g \dot{\phi_0}}},
 \end{equation}
where $\alpha_k={\rm Arg}\left(\Gamma\left(1/2+i\frac{-c_{\chi}^2 k^2 H^2 \tau_0^2}
{2 g \dot{\phi_0}}\right)\right)+\frac{-c_{\chi}^2 k^2 H^2 \tau_0^2}{2 g \dot{\phi_0}}
(1-\log{\frac{-c_{\chi}^2 k^2 H^2 \tau_0^2}{2 g \dot{\phi_0}}})$.

 With these initial conditions, we will now work in the non-relativistic
 limit as the Bogolyubov coefficients contain exponential momentum suppression, for which
 $\omega(|k-p|)-\omega(p)=0$ and $\omega^2=\frac{g^2 \dot{\phi_0}^2}{H^4 \tau^2}\left[\ln{\left(\frac{\tau_0}{\tau}\right)}\right]^2$.

Consequently, the two-point correlation function looks
\begin{multline}
\left\langle \gamma_{ij}(k,\tau)\gamma^{ij}(k',\tau)\right\rangle=\frac{\alpha_1^2}
{(\alpha_1+\alpha_2)^2}\frac{\delta(k+k')}{8 \pi^3 M_p^4}\int d^3p \left(p^2-\frac{\bf{p}.\bf{k}}{k^2}\right)^2 \\
\times  \int \frac{d\tau'}{a(\tau')^2}\frac{G_K(\tau,\tau')}{\sqrt{\omega_p(\tau')\omega_{k-p}(\tau')}}
\int \frac{d\tau''}{a(\tau'')^2}\frac{G_K(\tau,\tau'')}{\sqrt{\omega_p(\tau'')\omega_{k-p}(\tau'')}}\\
\times (2|\beta(p)|^4+2 |\alpha(p)|^2 |\beta(p)|^2).
\end{multline}



The $\tau \rightarrow 0$ limit of the above  Green's function is given by, 
$G_k(0,\tau')= \frac{c_{\gamma} k \tau' \cos(c_{\gamma} k \tau')-\sin(c_{\gamma} k \tau')}{c_{\gamma}^3 k^3 \tau'^2}$.
Hence, upon performing the $p$ and $\tau$ integration we get,
\begin{multline}\label{eq8}
\left\langle \gamma_{ij}(k,\tau)\gamma^{ij}(k',\tau)\right\rangle=
\frac{\delta(k+k')}{4 \pi^5 M_p^4} \frac{H}{c_{\gamma}^6 k^6 c_{\chi}^3}
\frac{(g\dot{\phi_0})^{3/2}}{\tau_0^3}\left(1+\frac{1}{4\sqrt{2}}\right)\\
\times \left(c_{\gamma}k\tau_0 \cos(c_{\gamma}k\tau_0)-\sin(c_{\gamma}k\tau_0)\right)^2
\left(\ln \frac{\sqrt{g\dot{\phi}_0}}{H}\right)^2.
\end{multline}
The role of non-trivial propagation speed $c_{\gamma}$ and $c_{\chi}$ are now crystal-clear
 from \eqref{eq8}. They can be used to tune the signal strength of the two-point function. For example, 
 it can be enhanced in the limit 
$c_{\gamma} \rightarrow 0$ or $c_{\chi}\rightarrow 0$ or $c_{\gamma},c_{\chi}\rightarrow0$. 
So, it is expected that they will play crucial role in determining the signal strength of three-point correlation functions as well. 
However, we will concentrate on this in the next section.

The total power spectrum for tensor modes reads
\begin{multline}\label{eq9}
P_{T}(k)=\frac{2H^2}{M_P^2 c_{\gamma}\pi^2}
\left[1+\frac{H^2}{M_p^2\pi^3 c_{\gamma}^5  c_{\chi}^3}\frac{(g\dot{\phi}_0)^{3/2}}{H^3}
\left(1+\frac{1}{4\sqrt{2}}\right)\right. \\ 
\times \left. \frac{\left(c_{\gamma}k\tau_0 \cos(c_{\gamma}k\tau_0)-\sin(c_{\gamma}k\tau_0)\right)^2}
{k^3 \tau_0^3} \left(\ln \frac{\sqrt{g\dot{\phi}_0}}{H}\right)^2 \right].
\end{multline}

It can be verified that the function 
$\frac{\left(c_{\gamma}k\tau_0 \cos(c_{\gamma}k\tau_0)-\sin(c_{\gamma}k\tau_0)\right)^2}{c_{\gamma}^3k^3 \tau_0^3}$ 
gets maximum value at $c_{\gamma}k\tau_0=2.46$.  In order to compare with the existing results in the literature, 
we take the same representative values for the parameter as in \cite{Cook}:
$g=1$, $H=10^{13} {\rm GeV}/c^2$, $M_p=2.48\times10^{18}{\rm  GeV}/c^2$ and $\dot{\phi}_0=\sqrt{2 \epsilon}H M_p$
 where, $\epsilon=0.005$. As a result, the tensor power spectrum becomes
\begin{equation}
\label{eq10}
P_{T}(k)=\frac{2H^2}{M_P^2 c_{\gamma}\pi^2}\left[1+6.75\times 10^{-6}\frac{1}{c_{\gamma}^2 c_{\chi}^3}\right].
\end{equation}

In the existing literature (e.g., \cite{Cook}), the second term in the parenthesis was generically small. However, in the present analysis,  it
can be significantly large  due to nontrivial speed of propagation. 
For example, if the second term is of order of one, the signal strength of two point correlation
function of PGW due to
(p)reheating can be of the same order of the vacuum contribution. 
Fig \ref{fig1}  demonstrates the comparative values of the two speed of propagation
 in order to achieve this.
\begin{figure} \label{fig1}
\includegraphics[width=7cm,height=12cm,keepaspectratio]{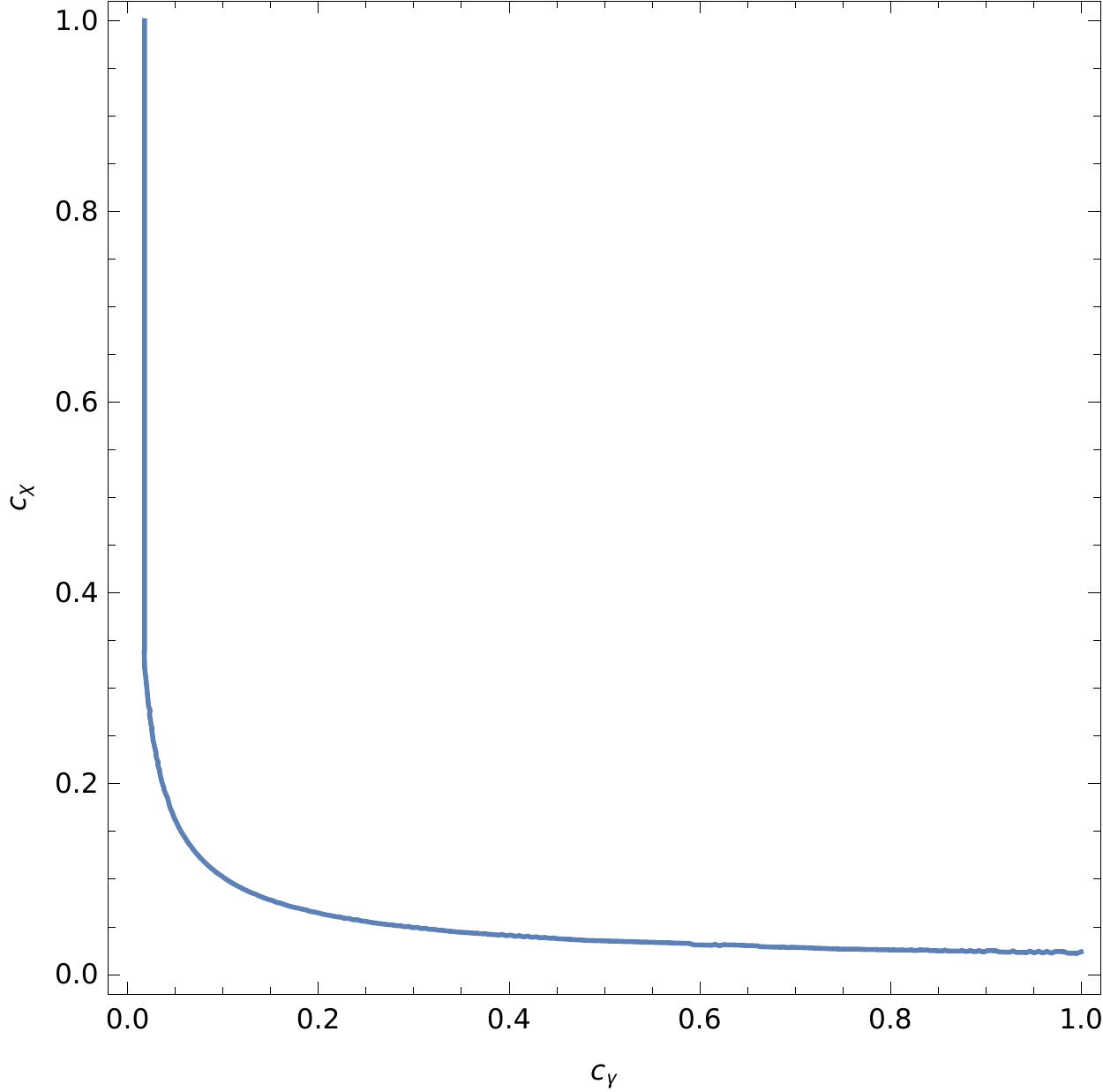}
	\caption{The correlation between $c_{\gamma}$ and $c_{\chi}$ for large contribution
	 of reheating sourced two point correlation function}
	\label{fig1}
\end{figure}
Let us explain it with a particular example. If we take a representative value for the tensor-to-scalar ratio as 
$r\approx 0.06$ that is close to the upper bound set by the latest Planck 2018 data \cite{Akrami:2018odb}, then
for $c_{\gamma}=1, ~c_{\chi}\approx 0.02$ the second term will be $\mathcal{O}(1)$. However, the signal strength of two point correlation function due to (p)reheating particles 
can be much larger than the signal strength due to vacuum fluctuations if  
$c_{\gamma}$ and $c_{\chi}$ become smaller than the above mentioned limit. Also we have noted
earlier that the signal strength gets maximum contribution for $c_{\gamma}k \tau_0=2.46$, so the
peak frequency of the signal will be dependent on $c_{\gamma}$. The peak frequency will be higher 
for a smaller $c_{\gamma}$. So the detectability of the signal is dependent on the EFT parameters and 
as explained above there lies a region in the parameter space where the signal strength becomes 
strong with peak frequency determined by $c_{\gamma}$.
This can be of interest for the upcoming gravitational wave (GW) missions such as the Einstein telescope 
\cite{Maggiore:2019uih}
which will operate in the high frequency limit where the GW signal strength produced from (p)reheating
gets peaked.

The reason for the enhancement of the signal is that for $c_{\chi}<1$ the resonance 
band become broadened and
there is an enhancement in particle production as discussed in  \cite{Giblin:2017qjp}.
On the other hand according to \cite{Karouby:2011xs} small propagation  speed  of tensor 
fluctuation is also responsible for large signal because non canonical inflationary case
is responsible for a saw-tooth like profile of inflaton which moves the system to broad
parametric resonance and significant particle production occurs.
Note that in the above analysis we did not consider the non-adiabatic scenario as it is shown in \cite{Cook}
that this regime produces same result as the adiabatic regime.


\section{Three-point correlation function}  \label{sec:3pt}

Having convinced ourselves about the role of the non-trivial propagation speed on the  signal strength, let us now move forward
to calculate the three-point function for (p)reheating-sourced gravitational waves.
The expression for three-point function is given by
\begin{multline}
\langle \gamma^{s_1}(k_1)\gamma^{s_2}(k_2)\gamma^{s_3}(k_3)\rangle=
 \left(\frac{-2\alpha_1}{2\pi^2 M_p^2}\right)^3 
 \int \frac{d\tau_1 d\tau_2 d\tau_3}{a(\tau_1)^2a(\tau_2)^2a(\tau_3)^2}\\
\times e^{s_1}_{i_1 j_1}e^{s_2}_{i_2 j_2}e^{s_3}_{i_3 j_3} \Pi_{i_1 j_1}^{ab}(k_1)\Pi_{i_2 j_2}^{cd}(k_2)\Pi_{i_3 j_3}^{ef}(k_3)
p_{1a}p_{1b}p_{2c}p_{2d}p_{3e}p_{3f}\\
\times \langle\chi(p_1,\tau_1)\chi(k_1-p_1,\tau_1)\chi(p_2,\tau_2)\chi(k_2-p_2,\tau_2)\chi(p_3,\tau_3)\chi(k_3-p_3,\tau_3)\rangle,
\end{multline}
where $s_i$ are helicity indices and $e^{s_i}_{ij}$ are polarization tensors.
 To fix the representation of polarization tensors we take a particular $k_i$
  basis and consider that this basis is lying on $(x,y)$ plane.
 In doing so   we will not lose any generality because of the momentum conserving $\delta$ function.
In what follows we will  choose the  representation adapted in \cite{Soda:2011am} :
$k_1=k_1(1,0,0)$, $k_2=k_2(\cos \theta_1, \sin \theta_1,0)$, $k_3=k_3(\cos \theta_2, \sin \theta_2,0)$ where\\
$\cos \theta_1=\frac{k_3^2-k_1^2-k_2^2}{2k_1 k_2}$,\\
 $\sin \theta_1=\frac{\sqrt{2k_1^2k_2^2+2k_2^2k_3^2+2k_1^2k_3^2-k_1^4-k_2^4-k_3^4}}{2k_1 k_2}$,\\
  $\cos \theta_2=\frac{k_2^2-k_1^2-k_3^2}{2k_1 k_3}$,\\
    $\sin \theta_2=-\frac{\sqrt{2k_1^2k_2^2+2k_2^2k_3^2+2k_1^2k_3^2-k_1^4-k_2^4-k_3^4}}{2k_1 k_3}$.

With this choice the polarization tensors can be written as,
\begin{equation}
e^{s_1}(k_1)=
\begin{pmatrix}
0 & 0 & 0 \\
0 & 1 & i s_1\\
0 & i s_1 & -1
\end{pmatrix},
\end{equation}
 
 \begin{equation}
 e^{s_2}(k_2)=
 \begin{pmatrix}
 \sin^2\theta_1 & -\sin\theta_1 \cos\theta_1 & -i s_2 \sin\theta_1\\
 -\sin\theta_1 \cos\theta_1 & \cos^2 \theta_1 & i s_2 \cos\theta_1\\
-i s_2 \sin\theta_1 & is_2 \cos\theta_1 & -1
  \end{pmatrix},
 \end{equation}
 
  \begin{equation}
 e^{s_2}(k_3)=
 \begin{pmatrix}
 \sin^2\theta_2 & -\sin\theta_2 \cos\theta_2 & -i s_2 \sin\theta_2\\
 -\sin\theta_2 \cos\theta_2 & \cos^2 \theta_2 & i s_2 \cos\theta_2\\
-i s_2 \sin\theta_2 & is_2 \cos\theta_2 & -1
  \end{pmatrix}.
 \end{equation}

 Consequently, the total three-point function gives us,
\begin{multline}\label{eqtot}
\langle \gamma^{s_1}(k_1)\gamma^{s_2}(k_2)\gamma^{s_3}(k_3)\rangle_{\rm total}= 
\langle \gamma^{s_1}(k_1)\gamma^{s_2}(k_2)\gamma^{s_3}(k_3)\rangle_{\rm vac}\\ + 
\langle \gamma^{s_1}(k_1)\gamma^{s_2}(k_2)\gamma^{s_3}(k_3)\rangle_{\rm so},
\end{multline} 
 where the subscripts "vac" and "so" stand for "vacuum" and  "source" (here, (p)reheating) respectively and these abbreviations would be used 
 in the rest of the article.
 
As already mentioned, the vacuum solution has been explored at length in a previous article by the present authors  \cite{Naskar:2018rmu} and is given as,
\begin{multline}\label{eq14}
\langle \gamma^{s_1}(k_1)\gamma^{s_2}(k_2)\gamma^{s_3}(k_3)\rangle_{\rm vac}=
(2 \pi)^3 \delta^{(3)}({k_1}+{k_2}+{k_3})\\
\times F(s_1 k_1,s_2 k_2,s_3 k_3) \\
\times
\left(\frac{64 H^4}{c_{\gamma}^2 M_{pl}^4}\frac{A(k_1,k_2,k_3)(s_1  k_1+s_2  k_2+s_3 k_3)^2}{k_1^3 k_2^3 k_3^3}\right. \\
\left. + \frac{4 \bar{M}_9 H^5}{M_{pl}^6}\frac{1}{k_1 k_2 k_3} \frac{1}{(k_1+ k_2+ k_3)^3}\right),
\end{multline}
where
$A(k_1,k_2,k_3)=\frac{K}{16}\left(1-\frac{1}{k^3} \sum_{i \neq j} k_i^2 k_j -\frac{4 k_1 k_2 k_3}{K^3}  \right)$
with ${K}=k_1+k_2+k_3$, and
$F(x,y,z)=-\frac{1}{64 x^2 y^2 z^2}(x+y+z)^3 (x+y-z) (x-y+z) (y+z-x).$

We will calculate the contribution from source term here.
 In evaluating the three-point function, we will use the same approximation of adiabatic regime as in the case of two-point function.
By employing this approximation, 
the source part of the three-point function takes the form
 \begin{multline}\label{eq30}
\langle \gamma^{s_1}(k_1)\gamma^{s_2}(k_2)\gamma^{s_3}(k_3)\rangle_{\rm so}=
-\left(\frac{2}{(2\pi M_p)^2}\right)^3 
\frac{\alpha_1^3}{(\alpha_1+\alpha_2)^3}\\
\times \frac{H^{12} \tau_0^6}{g^3\dot{\phi}_0^3k_1^3k_2^3k_3^3c_{\gamma}^9}
 \left(\ln \frac{\sqrt{g\dot{\phi}_0}}{H}\right)^3(\mathcal{A}_k+\mathcal{B}_k)\\
 \times  \prod_{i=1}^3 \left(c_{\gamma}k_i\tau_0 
  \cos(c_{\gamma}k_i\tau_0)-\sin(c_{\gamma}k_i\tau_0)\right)
   ,
 \end{multline}
 where the terms
$ \mathcal{A}_k$  and $\mathcal{B}_k$  have very tedious expressions. For completeness, we summaries them below:

\begin{multline}   
 \mathcal{A}_k=\frac{(g\dot{\phi}_0)^{\frac{7}{2}}}{124416 c_{\chi}^9H^9\pi^3\tau_0^9} 
 \frac{\left(k_1^4+(k_2^2-k_3^2)^2-2k_1^2(k_2^2+k_3^2)\right)}{k_1^2 k_2^2 k_3^2}\\
\times  \left(-3(81\sqrt{2}+16\sqrt{3})\right)\pi \tau_0^2 H^2 c_{\chi}^2 \\
\times \left\lbrace k_1^4+k_1^2(6k_2^2-2k_3^2)+(k_2^2-k_3^2)^2+4k_1^3 k_2 s_1 s_2 \right. \\
 \left. + 4 k_1 k_2 (k_2^2-k_3^2)^2 s_1 s_2 \right\rbrace+5~perms,
 \end{multline}
 
 \begin{multline}
\mathcal{B}_k= \frac{(g\dot{\phi}_0)^{\frac{7}{2}}}{124416 c_{\chi}^9H^9\pi^3\tau_0^9}
\frac{\left(k_1^4+(k_2^2-k_3^2)^2-2k_1^2(k_2^2+k_3^2)\right)}{k_1^2 k_2^2 k_3^2}g\dot{\phi}\\
\times 2(243 \sqrt{2}+32 \sqrt{3})  (k_1^2+k_2^2+k_3^2+2(2k_1 k_2 s_1 s_2+2k_1 k_3 s_1 s_3+2k_2 k_3 s_2 s_3)).
 \end{multline} 
Note that $\mathcal{B}_k$ is the sum of all six permutations.
 
  As mentioned, the resulting three-point function \eqref{eqtot} is the sumtotal of \eqref{eq14} and \eqref{eq30}.

 Let us now critically investigate for the results thus obtained. 
 To do so, we will have the following observations.
 First, from the expression of $ \mathcal{A}_k$  and $\mathcal{B}_k$ we can see that
 they can be written as,
 \begin{eqnarray}
  \mathcal{A}_k= C ( c_{\chi}^2 f(k)),\\
   \mathcal{B}_k=C  (g\dot{\phi}g(k)).
 \end{eqnarray}
 Where $C=\frac{1}{124416 c_{\chi}^8H^8\pi^3\tau_0^8}\sqrt{\frac{g\dot{\phi}}{c_{\chi}^2H^2\tau_0^2}}(g\dot{\phi})^3$
 and  $f(k)$ and $g(k)$ encodes all the momentum dependence and relevant prefactors.
 It is evident from the above expression that for a small $c_{\chi}$ we can neglect $\mathcal{A}_k$
 and only $\mathcal{B}_k$ contributes to the three point function.
Secondly, the term 
  $\left(c_{\gamma}k_i\tau_0 \cos(c_{\gamma}k_i\tau_0)-\sin(c_{\gamma}k_i\tau_0)\right)$ 
  can be expanded for small $c_{\gamma}$ and upto third order in $c_{\gamma}$ can be written as, $(c_{\gamma}k_i\tau_0)^3$.
  In order to extract out the momentum dependence of the bispectra from complicated functional form of 
  $\mathcal{B}_k$ we are working in a limit where we can keep up to $c_{\gamma}^3$ term and 
   can neglect $c_{\chi}^2$ term.

\begin{figure}
	\includegraphics[width=9cm,height=14cm,keepaspectratio]{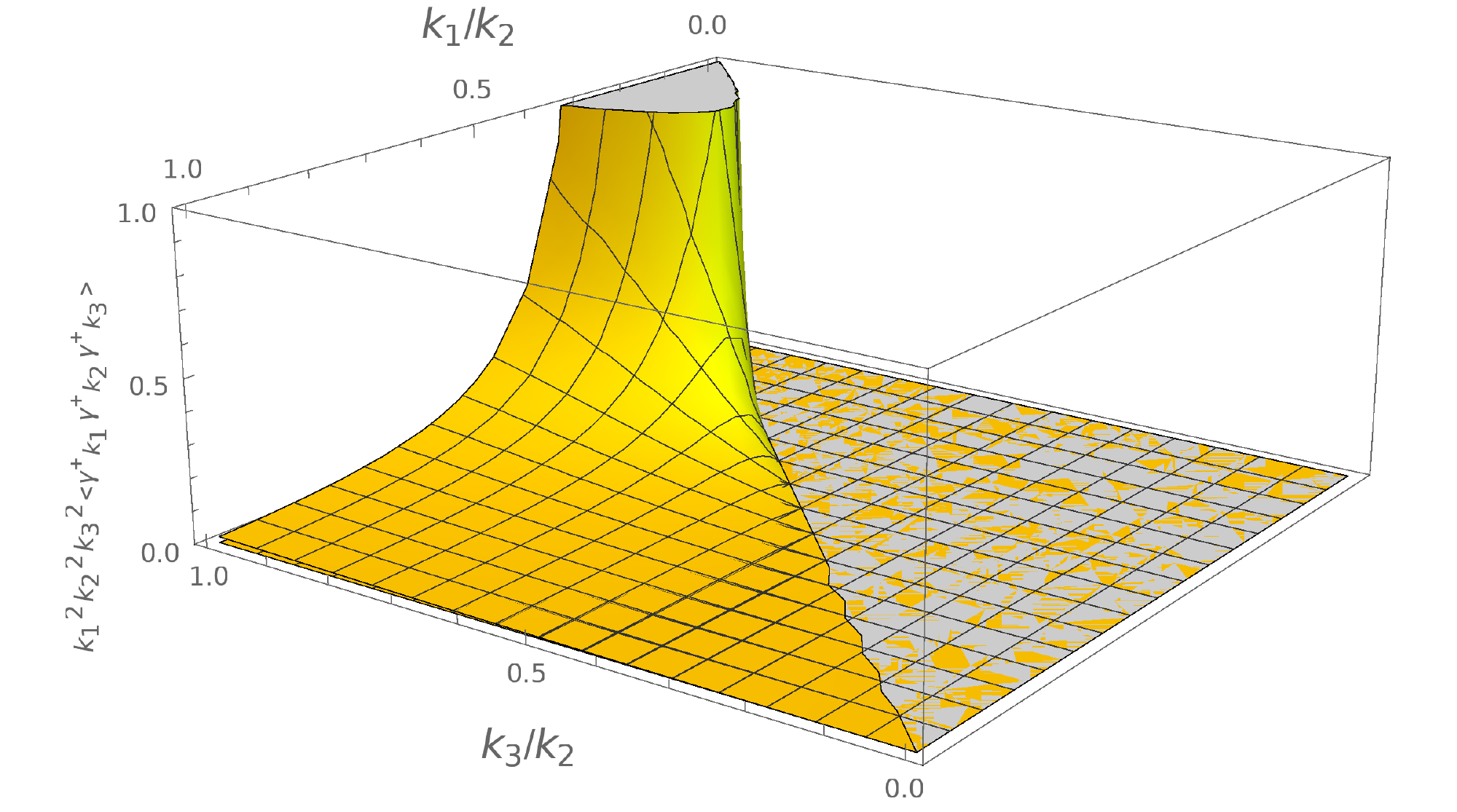}
	\caption{The bispectra is plotted as a function of $\frac{k_1}{k_2}$ and $\frac{k_3}{k_2}$}
	\label{fig2}
\end{figure}

   The resultant contributions have been pictorially depicted in
   Fig \ref{fig2}. The figure shows the momentum dependence 
  of  the bispectra  as a function of $\frac{k_1}{k_2}$ and $\frac{k_3}{k_2}$.
  The essential conclusion that  can be readily obtained from the above figure is that
  for $\frac{k_1}{k_2} \rightarrow 0$ and
  $0.755< \frac{k_3}{k_2}\leq 1$ we get large amplitude for the bispectra. This shows that 
  intermediate momentum configurations other than squeezed limit and equilateral limit
  can contribute significantly to the signal.
  Also we get positive contribution for squeezed and equilateral limit and much larger 
  amplitude for the bispectra which cannot be achieved in case of vacuum.
  This was the primary goal of the present article.
  We shall elaborate more on this in the following section.
  
\section{Estimation of $f_{NL}$} \label{sec:fnl}

We are now in a position to calculate the expressions for the nonlinearity parameter $f_{NL}$.
In what follows we shall make use of the same definition of  the non-linearity parameter  as adopted  in  \cite{Naskar:2018rmu}, 
namely,
$\frac{6}{5} f_{NL}=\frac{\langle \gamma \gamma \gamma \rangle}{P_{\zeta}(k_1)+P_{\zeta}(k_2)+P_{\zeta}(k_3)}$, where $P_{\zeta}(k)$ is the scalar powerspectrum and can be written as, \\
$P_{\zeta}(k)=\frac{2 \pi}{k^3} \frac{H^2}{8 \pi M_{pl}^2 c_s} \left(\frac{k}{k_*}\right)^{(n_s-1)}$
with $n_s$ and $c_s$ being the spectral tilt and sound speed of scalar perturbations respectively.
Also,  the tensor modes generated due to vacuum fluctuation would in any case be small, the templates for which have
already been proposed in  the previous article \cite{Naskar:2018rmu}. Hence, in this article 
we would be interested only about the three-point function due to source term 
$\langle \gamma^{s_1}(k_1)\gamma^{s_2}(k_2)\gamma^{s_3}(k_3)\rangle_{\rm so}$
in formulating the templates. 
As has been pointed out, we are interested about any significant enhancement of signal. 
Hence, we would consider the scenario where the  three-point function due to source term 
would have been dominant contribution to 
$\langle \gamma^{s_1}(k_1)\gamma^{s_2}(k_2)\gamma^{s_3}(k_3)\rangle_{\rm total}$ in Eq \eqref{eqtot}
and would investigate if this is achievable with the parameters under consideration.

Like the vacuum solution, in the case of equilateral limit $k_1=k_2=k_3$ we have two independent non-linearity parameters. 
They are  given by

\begin{multline}\label{eq33}
f_{NL}^{+++,eq}=f_{NL}^{---,eq}=
\frac{1945.07  g \dot{\phi}_0 \left(\ln \frac{\sqrt{g\dot{\phi}_0}}{H} \right)^3
}{M_p^2 c_{\gamma}^7 c_{\chi}^3k_1^3 \tau_0^3}\\
\times \left(c_{\gamma} k_1 \tau_0 \cos(c_{\gamma} k_1 \tau_0)-\sin(c_{\gamma} k_1 \tau_0)\right)^3\\
\times \left(\frac{c_s \epsilon}{c_{\gamma}}\right)^2
 \frac{\sqrt{g\dot{\phi}_0}}{H} \left(k_1/k_*\right)^{-2(n_s-1)},
\end{multline}

\begin{multline}
f_{NL}^{+-+,eq}=f_{NL}^{++-,eq}=f_{Nl}^{-+-,eq}\\
=f_{NL}^{--+,eq}=f_{NL}^{+--,eq}=f_{NL}^{-++,eq}=\\
\frac{216.12  g \dot{\phi}_0 \left(\ln \frac{\sqrt{g\dot{\phi}_0}}{H}\right)^3
(c_{\gamma} k_1 \tau_0 \cos(c_{\gamma} k_1 \tau_0)-\sin(c_{\gamma} k_1 \tau_0))^3}{M_p^2 c_{\gamma}^7c_{\chi}^3 k_1^3 \tau_0^3}\\
\times \left(\frac{c_s \epsilon}{c_{\gamma}}\right)^2
\frac{\sqrt{g\dot{\phi}_0}}{H} \left(k_1/k_*\right)^{-2(n_s-1)}.
\end{multline}



Consequently, for the squeezed limit, we get the following non-linearity parameters

\begin{multline}
f_{NL}^{+++,sq}=f_{NL}^{---,sq}=f_{NL}^{+--,sq}=f_{NL}^{-++,sq}\underset{k_1 \rightarrow 0}{=}\\
\frac{3457.89  g \dot{\phi} \left(\ln \frac{\sqrt{g\dot{\phi}}}{H}\right)^3 
}{M_p^2 c_{\gamma}^7 c_{\chi}^3k_2^3 \tau_0^3}\prod_{i=1}^3 
\left(c_{\gamma}k_i\tau_0 \cos(c_{\gamma}k_i\tau_0)-\sin(c_{\gamma}k_i\tau_0)\right)\\
\times \left(\frac{c_s \epsilon}{c_{\gamma}}\right)^2
\frac{\sqrt{g\dot{\phi}}}{H} \left(k_2/k_*\right)^{-2(n_s-1)}.
\end{multline}


We can see from the above expressions of $f_{NL}$ that a small propagation speed  
of either tensor fluctuations or preheating particles can lead to a large amplitude 
for tensor bispectrum. The non-Gaussian signal produced from (p)reheating can not
be observed in CMB scales but can be observable in GW interferometers.
However current interferometers still do not probe the scales where the signal can 
be detectable. But we should note that as the signal can be large for  
parameter combination mentioned above, the next iterations of the interferometers which
can probe higher frequencies can have a chance to detect them. 
 

 Here we consider CMB constraints on squeezed limit and equilateral limit bispectra 
\cite{Shiraishi:2019yux,Shiraishi:2013wua,Ade} to show the difference in
magnitude of equilateral and squeezed limit and to demonstrate how the constraint 
on $c_{\chi}$ changes, though one should remember that CMB constraint may not be
applicable to the derived $f_{NL}$.
 As we have stated earlier, from (p)reheating the two point function is 
peaked at $c_{\gamma}k_{i}\tau_0=2.46$ and for $c_{\gamma}=1$ and 
$c_{\chi}=0.02$ the signal strength becomes of the same order of vacuum contribution.
For squeezed limit  $f_{NL}$ where one  momentum is smaller than the
other two momenta, 
 we consider that $\frac{k_{\rm large}}{k_{\rm small}}\approx 10$. 
The constraint on
squeezed limit from Planck is $290\pm180$ \cite{Ade}. Using
the above approximations and  the upper limit of 
observational value of $f_{NL}^{+++,sq}=470$ we get $c_{\chi}>0.2$. 
Using the new constraint on $c_{\chi}$ we can estimate the 
$f_{NL}^{+++,eq} =0.3\left(k_1/k_*\right)^{0.071}$. Here we have used the best fit value for 
$n_s=0.9645$ from Planck 2018 \cite{Akrami:2018odb}. From these estimations we can see that
for $c_{\gamma}=1$ and small $c_{\chi}$ squeezed limit bispectrum is much larger than 
equilateral limit for PGW produced from
(p)reheating. This nature is also visible in Fig \ref{fig2}, but there we used an approximation
such that we can keep terms upto $c_{\chi}^3$ and neglect terms proportional to $c_{\chi}$. 
So for small $c_{\chi}$ squeezed limit will always be larger than the equilateral limit independent of whether 
$c_{\gamma}$ is small or unity.

Of course, these numerical estimations are not too accurate 
as we have considered the coupling constant to be $\mathcal{O}(1)$ 
which may not be strictly valid.
Also one have to use the late time
GW detectors' constraint on $f_{NL}$ to analyze the scenario.
In this work we refrain from commenting about the detectability of the signal
by upcoming GW missions rather our target was to  
demonstrate that using EFT in inflation and (p)reheating, 
large signal for tenor non-Gaussianities can be produced due to 
the presence of non trivial propagation speed of $\chi$ particles and tensor modes.

The bottomline of the above analysis is that we can have an enhanced tensor non-Gaussian signal from (p)reheating with 
non-trivial propagation speed $c_\chi$.  
 Also, particle production from non-canonical inflation with $c_\gamma < 1$
can enhance the tensor non-Gaussian signal further. 
A rather conservative statement would be that, the non-Gaussian signal produced from
(p)reheating can fall well
within the reach of next generation GW missions. As mentioned earlier Einstein telescope
will operate on the relevant frequency range to detect preheating produced GW signal
\cite{Maggiore:2019uih}, and this non-trivial non-Gaussian property of PGW 
can be of relevance for this kind of
 observations.
 However, an actual comparison with the sensitivity 
of upcoming GW  missions can only confirm this. 

\section{Conclusion}
In this article we have presented a way to enhance the signal for tensor 
three-point function sourced by  (p)reheating. 
Our analysis is based on EFT of inflation and (p)reheating, so we were able to analyze a large 
class of models where the interaction between inflaton and (p)reheating particle is described by
the choice of the EFT parameter $\alpha_3$.
Using EFT we have been able to deal
with a non standard case for (p)reheating for which the propagation speed of (p)reheat particle $\chi$
 is different from unity. We have demonstrated that tuning this non-trivial propagation speed 
 of (p)reheating particles along with
 the propagation speed of tensor fluctuation one can actually enhance 
the signal of tensor non-Gaussianities which was not achievable in the vacuum as well as in the standard (p)reheating analysis. 
We have further been able to propose templates for the non-linearity parameter $f_{NL}$ 
for these class of models and found
that, like the source-free case, here also squeezed limit bispectrum is stronger than  equilateral limit.
As a result, 
possibility of detection in future mission of the squeezed limit is higher 
along with the momentum range described in Section IV. 
 An actual comparison with the sensitivity 
of upcoming GW missions is beyond the scope of present article. We hope to address this issue with forecasts on 
couple of next-generation surveys in near future.

\section*{Acknowledgments}

AN thanks Indian Statistical Institute, Kolkata for financial support through Senior Research Fellowship.

\end{document}